\documentclass[11pt]{article}
\usepackage{epsfig}
\def\beq{\begin{equation}}
\def\eeq{\end{equation}}
\def\bea{\begin{eqnarray}}
\def\eea{\end{eqnarray}}
\def\L{{\cal L}}
\def\p{\partial}
\def\ep{\epsilon}
\def\dt{{d \over dt}~}

\begin{document}
\rightline{KOBE-TH-98-05}

\vspace{.5cm}

\begin{center}
{\LARGE Phase transition of a scalar field theory \\
\vspace{.2cm} at high temperatures}

\vspace{.5cm}
Hidenori SONODA\footnote{E-mail: sonoda@phys.sci.kobe-u.ac.jp}\\
Physics Department, Kobe University, Kobe 657-8501, Japan

\vspace{.2cm} 
September 1998
\end{center}

\begin{abstract}
At high temperatures a four dimensional field theory is reduced to a
three dimensional field theory.  In this letter we consider the $\phi^4$
theory whose parameters are chosen so that a thermal phase transition
occurs at a high temperature.  Using the known properties of the three
dimensional theory, we derive a non-trivial correction to the
critical temperature.
\end{abstract}

At very high temperatures the thermal properties of a 3+1 dimensional
field theory are given by an effective three dimensional field
theory.\cite{braaten-nieto} In this letter we use an effective field
theory to discuss the thermal phase transition of the $\phi^4$ theory at
a high temperature.  The idea is simple: we choose the parameters of the
$\phi^4$ theory in such a way that the phase transition occurs at a high
temperature for which the three dimensional reduction is a good
approximation.  We can then use the known properties of the three
dimensional $\phi^4$ theory to discuss the physics near and at the
transition temperature.  In ref.~\cite{braaten-nieto} the main interest
was in computing the free energy density of the massless theory at high
temperatures, and the thermal phase transition of a massive theory was
not discussed.  The method of effective field theory has been used
extensively to discuss the thermal phase transitions of the electroweak
theory.\cite{shapo} Applied to the simpler $\phi^4$ theory, the method
can give non-trivial results more easily because the three dimensional
$\phi^4$ theory is much better understood than the gauge theories.

The four dimensional theory is defined by the lagrangian
\beq
\L = {1 \over 2} \p_\mu \phi \p_\mu \phi + {m^2 \over 2} \phi^2 +
{\lambda \over 4!} \phi^4 + {\rm counterterms}~,
\eeq
where the counterterms are chosen in the $\overline{\rm MS}$ scheme with
$\bar{\mu} = 1$.  We choose $m^2$ to be negative so that the $Z_2$
symmetry is broken spontaneously at zero temperature.  Then a continuous
transition occurs approximately at temperature $T_c \simeq \sqrt{- 24
m^2 \over \lambda}$. \cite{weinberg} The $Z_2$ symmetry is restored at
temperatures $T > T_c$.  Observe that we can choose $- m^2$ to be of
order $T_c$ and $\lambda$ of order ${1 \over T_c}$.  If we take $T_c$
large, i.e., the coupling $\lambda$ small, so that the physical mass
$m_{\rm ph}$ is much smaller than $T_c$, we can reduce the four
dimensional theory to an effective three dimensional theory, and we can
use the well-known results on the phase transition of the three
dimensional $\phi^4$ theory.  Near the transition, $m_{\rm ph}$ is
small, and the condition $m_{\rm ph} \ll T_c$ is bound to be satisfied.

The three-dimensional $\phi^4$ theory is defined by the lagrangian
\beq
\L_{eff} = {1 \over 2} \p_\mu \Phi \p_\mu \Phi + {m_3^2 \over 2} \Phi^2
+ {\lambda_3 \over 4!} \Phi^4 + {\delta m_3^2 \over 2} \Phi^2~,
\eeq
where the spatial dimension is $3-\ep$, and the renormalization is done
in the $\overline{\rm MS}$ scheme with
\beq
\delta m_3^2 = {1 \over 2 \ep} C \lambda_3^2~,
\eeq
and
\beq
C = - {1 \over 6(4\pi)^2}~.\label{C}
\eeq
The parameters $m_3^2$ and $\lambda_3$ are related to the temperature
$T$ and the parameters of the four dimensional theory as follows
\cite{braaten-nieto}\footnote{The calculations in
ref.~\cite{braaten-nieto} largely depend on the earlier calculations in
ref.~\cite{arnold-zhai}.}:
\bea
&&\lambda_3 =T \left( \lambda + {\rm O} (\lambda^2)\right) \label{lambda}\\
&&m_3^2 = m^2 \left(~1+ {\lambda \over (4 \pi)^2}
\left( \ln T + k \right) + {\rm O} (\lambda^2) \right)\label{mtwo} \\
&&\quad + ~{\lambda T^2 \over 24}
\left( 1 + {\lambda \over (4\pi)^2}
\left( - \ln T + k - 2 \Delta\right) 
+ {\rm O} (\lambda^2) \right)~, \nonumber
\eea
where the constants are given by 
\beq
k = \ln 4 \pi - \gamma~,\quad \Delta = \ln 4 \pi - 1 - 
{\zeta'(-1) \over \zeta (-1)} \simeq -.454~.\label{const}
\eeq
($\gamma \simeq .577$ is Euler's constant, and $\zeta (z)$ is Riemann's
zeta function.)  Note that with the choice
\beq
m^2 = {\rm O} (T)~,\quad \lambda = {\rm O} \left(T^{-1}\right)~, 
\eeq
the above approximations (tree for $\lambda_3$, one-loop for the $m^2$
term in $m_3^2$, and two-loop for the $T^2$ term) give all the
contributions which survive the limit $T \to \infty$: the higher order
corrections vanish as $T \to \infty$.

The parameters $\lambda_3$ and $m_3^2$ of the three dimensional theory
satisfy the following renormalization group (RG) equations:
\beq
\dt \lambda_3 = \lambda_3~,\quad
\dt m_3^2 = 2 m_3^2 + C \lambda_3^2~.\label{rg-three}
\eeq
Note that these are consistent with eqns.~(\ref{lambda}, \ref{mtwo}) and
the one-loop RG equations of the four-dimensional theory:
\beq
\dt \lambda \simeq - {3 \lambda^2 \over (4 \pi)^2}~,\quad
\dt m^2 \simeq \left(2 - {\lambda \over (4 \pi)^2} \right)
m^2~.\label{rg-four} 
\eeq
($\dt T = T$ by convention.)

Let us summarize what is known about the phase transition of the three
dimensional $\phi^4$ theory. (See \cite{parisi}, especially chapter 8.)
Given $\lambda_3$, the $Z_2$ symmetry is exact for $m_3^2 > m_{3c}^2$,
and it is spontaneously broken for $m_3^2 < m_{3c}^2$.  Whether the
symmetry is broken or not must be determined by a RG invariant
criterion.  Using $\lambda_3$ and $m_3^2$, we can construct only one
independent RG invariant which can be chosen as
\beq
R (m_3^2, \lambda_3) \equiv {m_3^2 - C \lambda_3^2 \ln \lambda_3 \over
\lambda_3^2}~,
\eeq
where the constant $C$ is given by eqn.~(\ref{C}).  It is trivial to
check the RG invariance of $R$ using eqns.~(\ref{rg-three}).  Using $R$
we can rephrase the criterion for the transition: the symmetry is exact
for $R > R_c$, and broken for $R < R_c$, where $R_c$ is a constant.
This implies that
\beq
m_{3c}^2 = \lambda_3^2 (R_c + C \ln \lambda_3)~.
\eeq
The constant $R_c$ has not been calculated analytically.

\vskip .2cm
\centerline{\epsfig{file=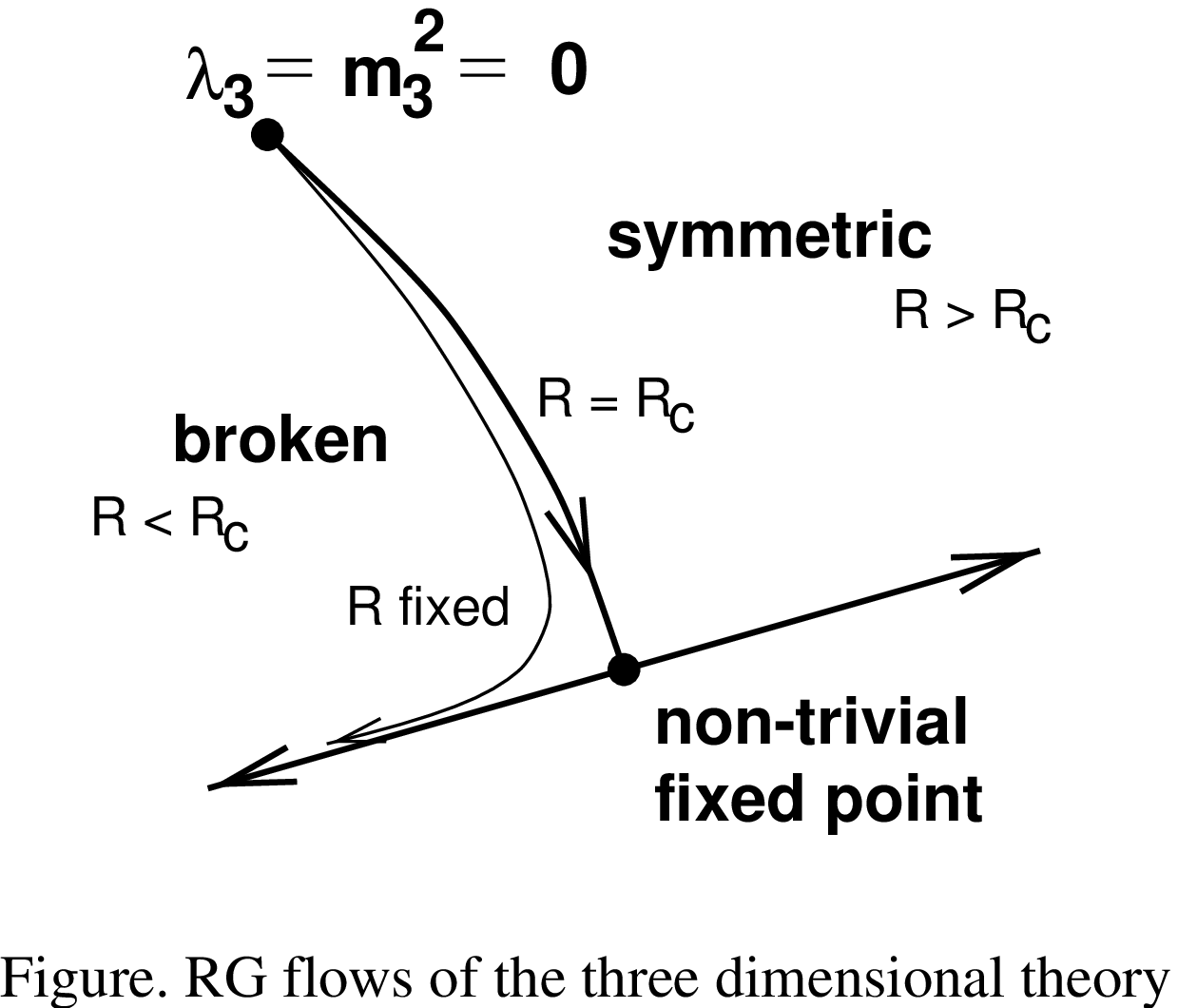, height=5cm}}
\vskip .2cm

The critical temperature $T_c$ of the four dimensional theory can be
obtained from the condition $m_3^2 = m_{3c}^2$.  Substituting
eqn.~(\ref{mtwo}) into this, we obtain the following expression for the
critical temperature
\beq 
T_c^2 = {- 24 m^2 \over \lambda}
\left[ 1 + {\lambda \over (4\pi)^2} \left(
- \ln (-24 m^2) - 3 \ln \lambda
+24(4\pi)^2 R_c + 2 \Delta \right) + {\rm O} (\lambda^2) \right]~,
\eeq
where $\Delta$ is given in eqns.~(\ref{const}).  We observe the
non-trivial logarithmic dependence on $\lambda$.\footnote{This is
similar to the appearance of the logarithm of $\lambda_3$ in the three
dimensional massless theory.\cite{symanzik}}

For completeness let us add a few comments about the critical behavior
of the theory.  (Again a good reference is ref.~\cite{parisi}.)  Clearly
the critical thermal behavior of the four dimensional theory is the same
as the critical behavior of the three dimensional theory.  In
particular, all the critical exponents are the same.  For example, near
$T=T_c$ the physical mass $m_{\rm ph}$ of the theory behaves as
\beq
m_{\rm ph}^2 \simeq {\rm const} \cdot \lambda_3^2 (R - R_c)^{2 \over y_E}
\simeq {\rm const} \cdot \lambda^{2-{2 \over y_E}} T_c^2
\left({T-T_c \over T_c}\right)^{2 \over y_E}~,
\eeq
where $y_E$ is the scale dimension of the relevant parameter at the
non-trivial fixed point.  The approximate value of $y_E$ has been
calculated by various methods: for example, the one-loop Callan-Symanzik
equation gives \cite{parisi}
\beq
y_E \simeq {5 \over 3}~.
\eeq
Similarly, at the critical temperature, the two-point thermal
correlation function (Matsubara function) of $\phi$ behaves as
\beq
\left\langle \phi (r) \phi (0) \right\rangle_{T=T_c} \simeq
T_c \left\langle \Phi (r) \Phi (0) \right\rangle_{m_3^2=m_{3,c}^2} 
\simeq {\rm const} \cdot {\lambda^{- \eta} T_c^{1-\eta} \over r^{1+\eta}}~,
\eeq
where the anomalous dimension $\eta$ is about $.05$.

How can we improve the high temperature approximation?  At higher orders
we must not only calculate the next loop order terms in
eqns.~(\ref{lambda},\ref{mtwo}), but also we must introduce
irrelevant\footnote{irrelevant with respect to the fixed point at
$\lambda_3 = m_3^2 = 0$ as opposed to the non-trivial infrared fixed
point} terms such as $\Phi^6$ (dimension three) and $\Phi^2 \p_\mu \Phi
\p_\mu \Phi$ (dimension four) in the three dimensional lagrangian.  The
calculations will be significantly more complicated.  A simple analysis
shows that the parameter of $\Phi^6$ is of order ${1 \over T^3}$, and
that of $\Phi^2 \p_\mu \Phi \p_\mu \Phi$ is of order ${1 \over T^2}$.
Therefore, we still do not need to introduce irrelevant terms at the
next order, but at the next next order we must introduce the term
$\Phi^2 \p_\mu \Phi \p_\mu \Phi$.

The above calculation of the critical temperature can be easily extended
to the four dimensional O(N) linear sigma model whose lagrangian is
given by
\beq
\L = {1 \over 2} \sum_{I=1}^N \p_\mu \phi^I \p_\mu \phi^I 
+ {m^2 \over 2} \sum_{I=1}^N \phi^I \phi^I
+ {\lambda \over 8}  \left( \sum_{I=1}^N \phi^I \phi^I \right)^2
+ {\rm counterterms}~.
\eeq
The critical temperature is obtained as
\bea
T_c^2 &=& {- 24 m^2 \over (N+2) \lambda} \Bigg[ ~1 +
{3 \lambda \over (4\pi)^2}
\Big\lbrace - \ln (- 24 m^2)\nonumber\\
&& - 3 \ln \left((N+2) \lambda\right)
+ 8 (N+2) (4\pi)^2 R_{N,c} + 2 \Delta ~\Big\rbrace 
+ {\rm O} (\lambda^2)~\Bigg],
\eea
where the constant $R_{N,c}$ is the value of the RG invariant 
\beq 
R_N (m_3^2, \lambda_3) \equiv {m_3^2 \over (N+2)^2 \lambda_3^2} -
C_N \ln \left((N+2)\lambda_3\right)
\eeq 
at the critical point ($C_N \equiv - {1\over 2(N+2) (4\pi)^2}$).  In the
large N limit we find $R_{N,c} = 1/(8 \pi)^2$.

\vskip .5cm
In this letter we have computed the critical temperature of the four
dimensional $\phi^4$ theory to the next leading order in the small
coupling constant $\lambda$ using the effective three dimensional
theory.

\end{document}